\documentclass[prd,twocolumn,amsmath,amssymb,floatfix,superscriptaddress,nofootinbib]{revtex4-1}
\usepackage{bm}
\usepackage{amsmath}
\usepackage{epsfig}
\usepackage{color}
\usepackage{natbib}
\usepackage{graphicx}
\usepackage{ifthen}
\usepackage{xstring}
\usepackage{graphicx}

\newcommand{\xef}{x_e^{\rm fid}}

\newcommand{\zmax}{z_{\rm max}}
\newcommand{\zmin}{z_{\rm min}}

\newcommand{\beq}{\begin{equation}}
\newcommand{\eeq}{\end{equation}}

\newcommand{\bea}{\begin{eqnarray}}
\newcommand{\eea}{\end{eqnarray}}

\definecolor{darkgreen}{cmyk}{0.85,0.2,1.00,0.2} 
\definecolor{purple}{cmyk}{0.5,1.0,0,0}

\definecolor{ultramarine}{rgb}{0.07, 0.04, 0.56}
\definecolor{cadmiumgreen}{rgb}{0.0, 0.42, 0.24}
\definecolor{indigo(dye)}{rgb}{0.0, 0.25, 0.42}
\usepackage[linktocpage=true]{hyperref}
\hypersetup{
colorlinks=true,
citecolor=ultramarine,
linkcolor=cadmiumgreen,
urlcolor=indigo(dye),
pdfauthor={},
pdftitle={},
pdfsubject={}
}

\begin{document}
\title{Complete reionization constraints from Planck 2015 polarization}

\author{Chen He Heinrich*}
\affiliation{Kavli Institute for Cosmological Physics,
Enrico Fermi Institute, University of Chicago, Illinois 60637}
\affiliation{Department of Physics,
 University of Chicago, Illinois 60637, USA}
 \thanks{Corresponding author: \href{mailto:chenhe@uchicago.edu}{chenhe@uchicago.edu}}
 
 \author{Vinicius Miranda}
\affiliation{Center for Particle Cosmology, Department of Physics and Astronomy, University of Pennsylvania, Philadelphia, Pennsylvania 19104, USA}

\author{Wayne Hu}
\affiliation{Kavli Institute for Cosmological Physics,
Enrico Fermi Institute, University of Chicago, Chicago Illinois 60637, USA}
\affiliation{Department of Astronomy \& Astrophysics,
 University of Chicago, Illinois 60637, USA}

\date{\today}

\begin{abstract}
We conduct an analysis of the Planck 2015 data that is complete in reionization observables 
from the large angle polarization $E$-mode spectrum
in the redshift range  $6 < z < 30$.   Based on 5 principal components, all of which are constrained by the data, this single analysis can be used to infer constraints on
any model for reionization in the same range; we develop an effective likelihood approach for applying these
constraints to models.   
By allowing for an arbitrary ionization history, this technique tests the robustness of 
inferences on the total optical depth from the usual steplike transition assumption, which is
important for the interpretation of many other cosmological parameters such as the dark energy 
and neutrino mass.   The Planck 2015 data not only allow a high redshift $z>15$ component to the optical depth but prefer it at the $2\sigma$ level.   This preference is associated with excess power in the multipole range $10 \lesssim \ell \lesssim 20$ and may  indicate high redshift ionization sources or unaccounted for systematics and foregrounds in the 2015 data. 
\end{abstract}

\maketitle

\section{Introduction} \label{sec:intro}

The epoch of reionization of the Universe remains one of the least well-understood aspects of the standard
model of cosmology (see e.g.~\cite{2016ASSL..423.....M}).   Yet its impact on the interpretation of its fundamental properties is comparatively large in this era of precision cosmology.
 In addition to the intrinsic astrophysical interest in ionization sources, reionization uncertainties
 impact cosmic microwave background (CMB) inferences on the initial power spectrum and hence also cosmic acceleration through the
  growth of structure \cite{Hu:2003pt}. In the future it will be one of the leading sources of error in the
  interpretation of neutrino mass measurements from gravitational lensing
  \cite{Smith:2006nk,Allison:2015qca}, the study of large scale anomalies in the CMB
  \cite{Mortonson:2009xk,Mortonson:2009qv}, and the inflationary consistency relation \cite{Mortonson:2007tb}. 
  
The standard approach to parametrizing the impact of reionization on the CMB is with the total Thomson
optical depth through the reionization epoch.  Although it is indeed the total optical
depth that is important for the interpretation of most other aspects of cosmology, it is usually
assumed that reionization occurs promptly in a steplike transition.   Interestingly, the central value of the inferred optical depth from this approach has been steadily drifting downwards from its first 
detection in WMAP1 \cite{Kogut:2003et}
to the current but still proprietary Planck 2016 High Frequency Instrument (HFI) results  \cite{Aghanim:2016yuo,Adam:2016hgk}.

Relaxing this sharp transition assumption
can in principle raise  the optical depth inference from the CMB  as well as change its implications for sources of high redshift ionization (e.g.~\cite{2012ApJ...756L..16A}).  In particular, the angular scale of the peak and the width of the reionization bump in the $E$-mode CMB
polarization power spectrum carries coarse grained information on the redshift dependence of
the ionization history.

There is an alternate, model independent approach introduced in Ref.~\cite{Hu:2003gh} that fully
addresses these concerns.
The impact of {\it any} ionization history on the large angle CMB polarization spectrum can be completely
characterized by a handful of reionization parameters constructed from the principal components (PCs) of the ionization history with respect to  the $E$-mode power spectrum.  This approach has the advantage over redshift binning alternatives of being observationally complete without introducing numerous highly correlated parameters
\cite{Lewis:2006ym}.  Conversely, it does not provide an accurate, local reconstruction for visualizing the
ionization history itself.  

This approach was implemented and tested on WMAP3  \cite{Mortonson:2007hq} and WMAP5 \cite{Mortonson:2008rx} power spectra which showed that those data allowed for, but did not particularly
favor, contributions to the optical depth from high redshift.   It was adopted in the Planck 2013 analysis but 
exclusively 
to test the impact of marginalizing the ionization history on inflationary parameters rather than drawing inferences
on reionization itself
\cite{Planck:2013jfk}.   The impact on massive neutrinos and gravitational wave inferences
was also examined in Ref.~\cite{Dai:2015dwa}. 

In this work, we analyze the public Planck 2015 data, including the Low Frequency Instrument (LFI) large angle
polarization power spectrum, using the observationally complete
PC basis.   In addition we further develop the technique as a method to probe reionization itself.
This development is timely as the technique should come 
 to its full fruition with the upcoming final release of Planck data which will be the definitive
 result on large angle polarization for years to come.
 
We demonstrate that the Planck 2015 data already have more information on the  ionization history than
just the total optical depth and provide  effective likelihood tools for interpreting
this information in any given model for reionization within the redshift range analyzed.  Tested here, these 
techniques can be straightforwardly applied to the final release when it becomes available.

We begin with a review of the approach itself in Sec. \ref{sec:PC}.  In Sec. \ref{sec:MCMC}, we analyze the Planck 2015 data and explore the origin of the additional information on
the high redshift ionization history.   We develop and test an effective likelihood approach
for utilizing our analysis to constrain the parameters of any given model of reionization from
$6 < z <30$
in Sec. \ref{sec:likelihood}.   We discuss these results in Sec. \ref{sec:discussion}.

\section{Complete Reionization Basis} 
\label{sec:PC}

We briefly summarize the principal component technique for the complete characterization
of reionization constraints from the large angle $C_\ell^{EE}$ polarization spectrum
as introduced in Ref.~\cite{Hu:2003gh} and implemented in \cite{Mortonson:2008rx,Mortonson:2007hq}.

We parametrize the ionization fraction relative to fully ionized hydrogen
$x_e(z)$ into its
principal components  with respect to the $E$-mode polarization of the 
CMB~\citep{Hu:2003gh}:
\begin{equation}
x_e(z)=\xef(z)+\sum_{a}m_{a}S_{a}(z).
\label{eq:mmutoxe}
\end{equation}
Here $m_a$ are the PC amplitudes and $S_{a}(z)$
are the eigenfunctions of the  Fisher information matrix for $x_e(z)$ in a given range  $z_{\rm min}<z<z_{\rm max}$ from cosmic variance limited $C_\ell^{EE}$ measurements,
and
$\xef(z)$ is the fiducial model around which the Fisher matrix is computed.  
In practice, we discretize the redshift space to $\delta z= 0.25$, well beyond the
resolution limit of CMB observables, and assume linear interpolation between points to form the
continuous functions $S_a(z)$.
The
components are rank ordered by their Fisher-estimated variances and  in practice
the first 5 components carry all the information in $C_\ell^{EE}$ to the cosmic variance limit
\cite{Hu:2003gh}.  In this work, we therefore truncate the PC expansion and retain 5 $m_a$
parameters to describe reionization.  We take $z_{\rm min}=6$ to be consistent with Ly$\alpha$
forest constraints (e.g. \cite{Becker:2015lua}) and $z_{\rm max}=30$.

In this truncated representation, the 5 PC decomposition is {\it not} complete in the ionization history itself.   
Instead it is a complete representation of the {\it observable} impact on $C_\ell^{EE}$ of
any given $x_e^{\rm true}(z)$ through its projection onto the 5 PC basis
\begin{equation}
m_{a}=
  \int _{\zmin}^{\zmax} dz\, \frac{S_{a}(z) [x_e^{\rm true}(z)-\xef(z)]}{\zmax-\zmin} .
\label{eq:xetommu}
\end{equation}
When reconstructed through Eq.~(\ref{eq:mmutoxe}) with 5 PCs, $x_e(z) \ne x_e^{\rm true}(z)$, even 
though it models the observables to high precision.   The PC analysis therefore
is a forward tool to infer constraints on all possible ionization histories between $z_{\rm min}<z<z_{\rm max}$ with a single analysis, not an inverse tool that reconstructs the ionization history.

We use a modified version of CAMB\footnote{CAMB: \url{http://camb.info}}~\cite{Lewis:1999bs, Howlett:2012mh} to compute the CMB power spectra given the ionization history.  
Because CAMB integrates the Boltzmann equation by parts, it requires a smooth ionization 
history for numerical stability, whereas the $S_a(z)$ are continuous but not smooth.    Consequently we smooth the ionization history in Eq.~(\ref{eq:mmutoxe}) with a 
Gaussian in $\mathrm{ln}(1+z)$ of width $\sigma_{\mathrm{ln}(1+z)} = 0.015$.   This does not affect our results in a statistically significant way.  
However for consistency,
when integrating the ionization history to form the cumulative
Thomson optical depth 
\begin{equation}
\tau(z,z_{\rm max}) = n_{\rm H}(0) \sigma_T \int_z^{z_{\rm max}} dz \frac{x_e(z) (1+z)^{2} }{H(z)} ,
\label{eq:cumtau}
\end{equation}
 we employ the smoothed $x_e$ which
formally has support beyond the bounds. 
 We include this small correction by integrating
slightly past $z_{\rm max}$ in practice.  Here $n_{\rm H}(0)$ is the hydrogen number density at $z=0$.

\begin{figure}[t]
          \includegraphics[width=0.95\linewidth]{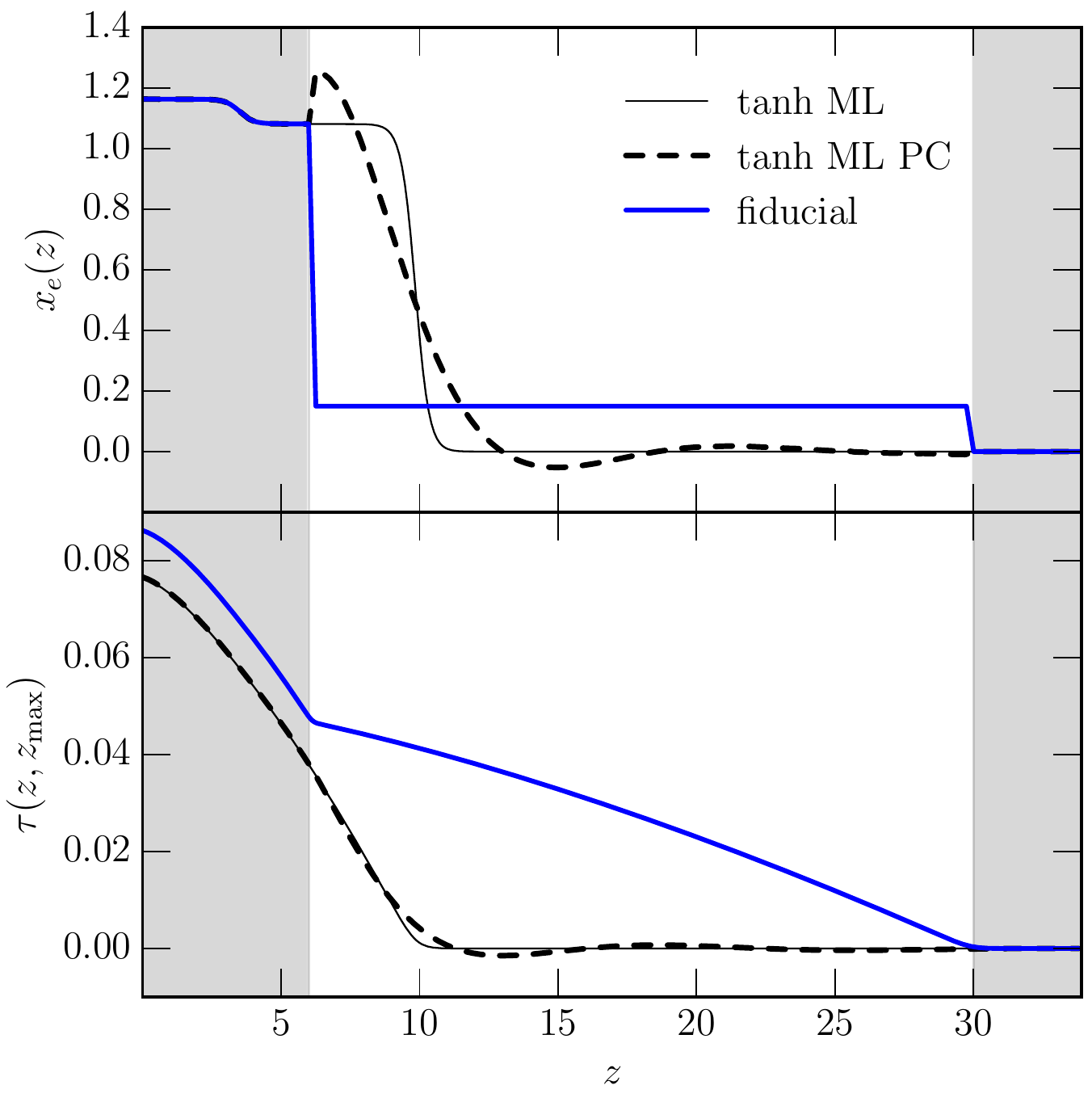} 
             \caption{\footnotesize Tanh reionization (black thin) vs~fiducial  model (blue thick) around which the PC decomposition (in the unshaded region) is centered. Top: Ionization fraction $x_e(z)$ as a function of redshift.  For the tanh model, despite modeling the
             polarization observables to high accuracy, the $x_e$ reconstruction from 5 PCs is poor (black dashed). 
             Bottom: 
             Cumulative optical depth from $z$ to $z_{\rm max}$.  
              The tanh model corresponds to the maximum likelihood in the MCMC chain of Sec. \ref{sec:MCMC}
              and with its sudden reionization accumulates its optical depth exclusively at low redshift in
              contrast with the fiducial model.  The 5 PC reconstruction is a better approximation for
              $\tau(z,z_{\rm max})$, especially for the difference between low and high $z$.}
            \label{fig:reion_models}   
\end{figure}

For the fiducial ionization history, we take $\xef=0.15$ for $6<z<30$ on the $\delta z=0.25$ spaced discrete points in the interval with linear interpolation in between,
 in order to let the PC amplitudes
fluctuate the ionization history both up and down without entering the unphysical region 
$x_e <0$,  where the number is chosen to give a reasonable $\tau(0,z_{\rm max})$.  It is important to note that the PCs allow
arbitrarily large deviations from the fiducial model, and so this choice does not bias
results.    For $z\ge 30$ we assume $x_e$ follows the ionization history from
recombination. For $z\le 6$ we  assume fully ionized hydrogen and singly ionized helium
 \begin{equation}
 x_e = 1+ f_{\rm He},
 \end{equation}
 for $z \gtrsim z_{\rm He}$ and doubly ionized helium
  \begin{equation}
x_e =  1+ 2f_{\rm He},
 \end{equation}
for $z \lesssim z_{\rm He}$,
where
\beq
f_{\rm He} = \frac{n_{\rm He}}{n_{\rm H}} = \frac{m_{\rm H}}{m_{\rm He}}\frac{Y_p}{1-Y_p}
\eeq
is the ratio of the helium to hydrogen number density.
We  take the helium mass fraction $Y_p$ to be consistent with big bang nucleosynthesis given the baryon density. 
  Following CAMB, we take this helium reionization transition to be mediated by
 a tanh function in redshift centered at  $z_{\rm He} = 3.5$ \cite{Becker:2010cu} with width $\Delta z=0.5$.
 \begin{figure}[t]
          \includegraphics[width=0.95\linewidth]{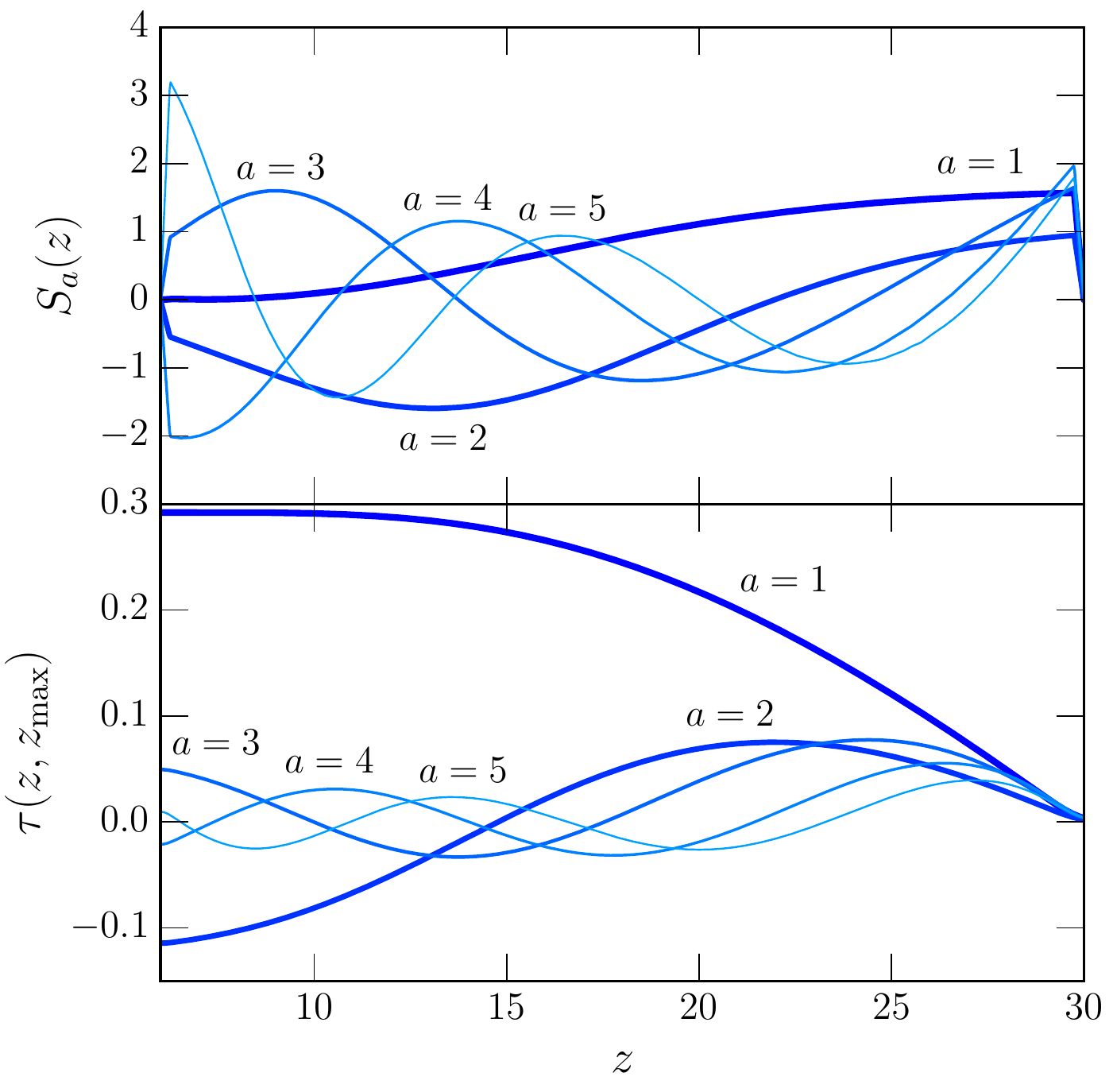} 
             \caption{\footnotesize First five PCs forming a complete set of polarization observables for the redshift range displayed.  Top: Ionization PCs in redshift.
             PCs are continuous functions with linear interpolation between discrete values at $\delta z=0.25$ with boundaries at $z_{\rm min}$ and $z_{\rm max}$ held fixed to zero. Bottom: Cumulative optical
             depth for unit amplitude modes as in Fig.~\ref{fig:reion_models}.  The first mode ($a=1$) mainly changes
             the total optical depth $\tau(0,z_{\rm max})$ from the fiducial model.  The second mode ($a=2$) adjusts the low and high redshift
             optical depth contributions whereas the higher modes make finer and less observable
             modifications to its redshift distribution.} 
            \label{fig:reion_basis}   
\end{figure}

In Fig.~\ref{fig:reion_models}, we show the fiducial ionization history (thick blue) and contrast it with
the standard approach of CAMB (thin black) that takes hydrogen and singly ionized helium reionization to be given by
the tanh form
 \begin{equation}
x_e^{\rm true}(z) = \frac{1+f_{\rm He}}{2}\left\{  1+ \tanh\left[ \frac{y(z_*)-y(z)}{\Delta y} \right] \right\},
 \label{eqn:tanh}
 \end{equation}
 with $y(z)=(1+z)^{3/2}$, $\Delta y=(3/2)(1+z)^{1/2}\Delta z$, and $\Delta z = 0.5$.  We take here $z_*= 9.85$, corresponding the chain maximum likelihood (ML) model ($\tau = 0.0765$) from \S \ref{sec:MCMC},  for illustrative purposes.   Projected onto 5 PCs and resummed into $x_e(z)$, Eq.~(\ref{eq:mmutoxe}) yields a poor reconstruction of the ionization history itself.
 Nonetheless as we shall see in Fig.~\ref{fig:clee}, the PC decomposition provides an 
 excellent representation of the polarization power spectrum.

 We also show in Fig.~\ref{fig:reion_models} the cumulative optical depth from $z$ to $z_{\rm max}$.    Although the
two models have comparable total optical depth $\tau(0,z_{\rm max})$ the fiducial model 
receives much of its contribution from high redshift.   Note also that although we do not 
allow for uncertainties in helium reionization, its entire impact is a small correction on an already small contribution to $\tau$.  Furthermore, although the reconstruction of $\tau(z,z_{\rm max})$ using 5 PCs 
is still imperfect (black dashed line), it is much better than $x_e(z)$ as it is more closely related to the polarization
observables.    The PC reconstruction smooths out sharp transitions in the cumulative
optical depth but gives an accurate representation of the high and low redshift contributions
of the model.   In the analysis below, in order to compare exactly the same statistic $\tau(z,z_{\rm max})$ between models, we always employ the PC reconstructed version.

In Fig.~\ref{fig:reion_basis} we show the 5 PC ionization functions $S_a(z)$ which allow
observationally complete variations around the fiducial model.   We also show  the cumulative
optical depth $\tau(z,z_{\rm max})$ for a unit amplitude $m_a$ in each mode.
The lowest-variance eigenmode $S_1$ adjusts the total optical depth, mainly from the
high redshift end.   The $S_2$ mode allows a redistribution of the optical depth between
high redshift and low redshift.   The higher modes allow finer adjustments in the redshift
distribution of the optical depth and carry very little total optical depth $\tau(0,z_{\rm max})$.

For the Planck data set, most of the information in the ionization history is carried by the
first two modes and therefore relates to the amount of high vs.~low redshift optical depth.   
We keep all 5 PCs  for completeness in representing the observable
impact of a given ionization history and to marginalize uncertainties that they introduce.

\begin{figure*}
          \includegraphics[width=0.75\linewidth]{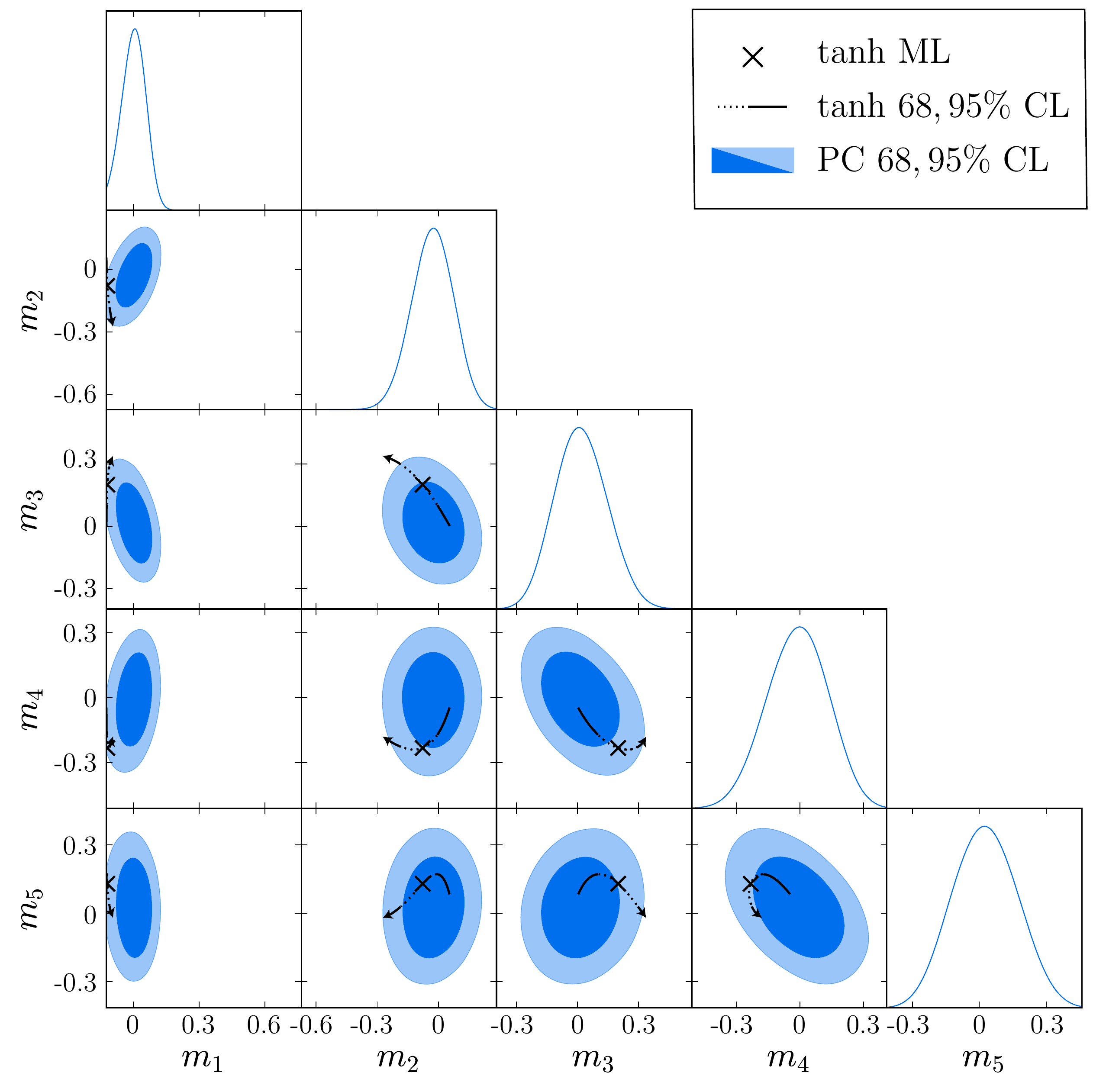}
             \caption{\footnotesize Reionization constraints from Planck 2015 data on PC amplitudes compared with those of  the tanh model projected onto PCs.  1D posterior
             probability distributions are shown on the upper diagonal and 2D 68\% and 95\% CL regions in the
              $m_a$-$m_b$ planes  in the lower triangle.   Tanh constraints 
                             are shown as trajectories in these
              planes (68\% CL dotted lines; 95\% CL solid lines) with the arrow pointed to higher $\tau$. 
              The  maximum likelihood ($\times$) and allowed region for the tanh model is disfavored in the observationally complete PC parameter space.  Box boundaries represent physicality priors on the
              ionization history.}
            \label{fig:mj_triangle_planck2015}   
\end{figure*}

In fact, PCs with no prior constraints on the mode amplitudes $m_a$
allow deviations that are unphysical $x_e<0$ and $x_e >x_e^{\rm max}$.  With a truncation at 
5 PCs, physicality cannot be strictly enforced since the missing eigenmodes, while irrelevant
for the observables, can restore physicality of a model.   We follow Ref.~\cite{Mortonson:2008rx} in placing necessary but not sufficient conditions for physicality  
\begin{equation}
\sum_{a=1}^5 m_a^2 \le (x_e^{\rm max}-x_e^{\rm fid})^2,
\end{equation}
where $x_e^{\rm fid}=0.15$ and $m_a^{-} \le m_a \le m_a^{+}$ with
\begin{equation}
m_a^{\pm} = \int_{z_{\rm min}}^{z_{\rm max} } dz \frac{S_a(z)[x_e^{\rm max} -2 x_e^{\rm fid}(z)]
\pm x_e^{\rm max} | S_a(z)|}{2(z_{\rm max}-z_{\rm min})}.
\label{eq:individualprior}
\end{equation}
In principle $x_e^{\rm max}=1+ 2 f_{\rm He}$ but in practice for compatibility with the ionization state  at $z_{\rm min}$ we
take $x_e^{\rm max} =1+f_{\rm He}$.   Since this prior is applied very conservatively as a necessary condition, we in fact still allow
ionization fractions out beyond the true maximum with this prior.

For analyzing any physical model  of hydrogen reionization, $x_e^{\rm true}(z)$, using our 5 PC decomposition from
Eq.~(\ref{eq:xetommu}), the priors are noninformative
as they are automatically satisfied.  We apply them mainly for visualizing whether the data 
 constrain the mode amplitudes significantly better than the physicality bounds.

\begin{figure}
          \includegraphics[width=0.90\linewidth]{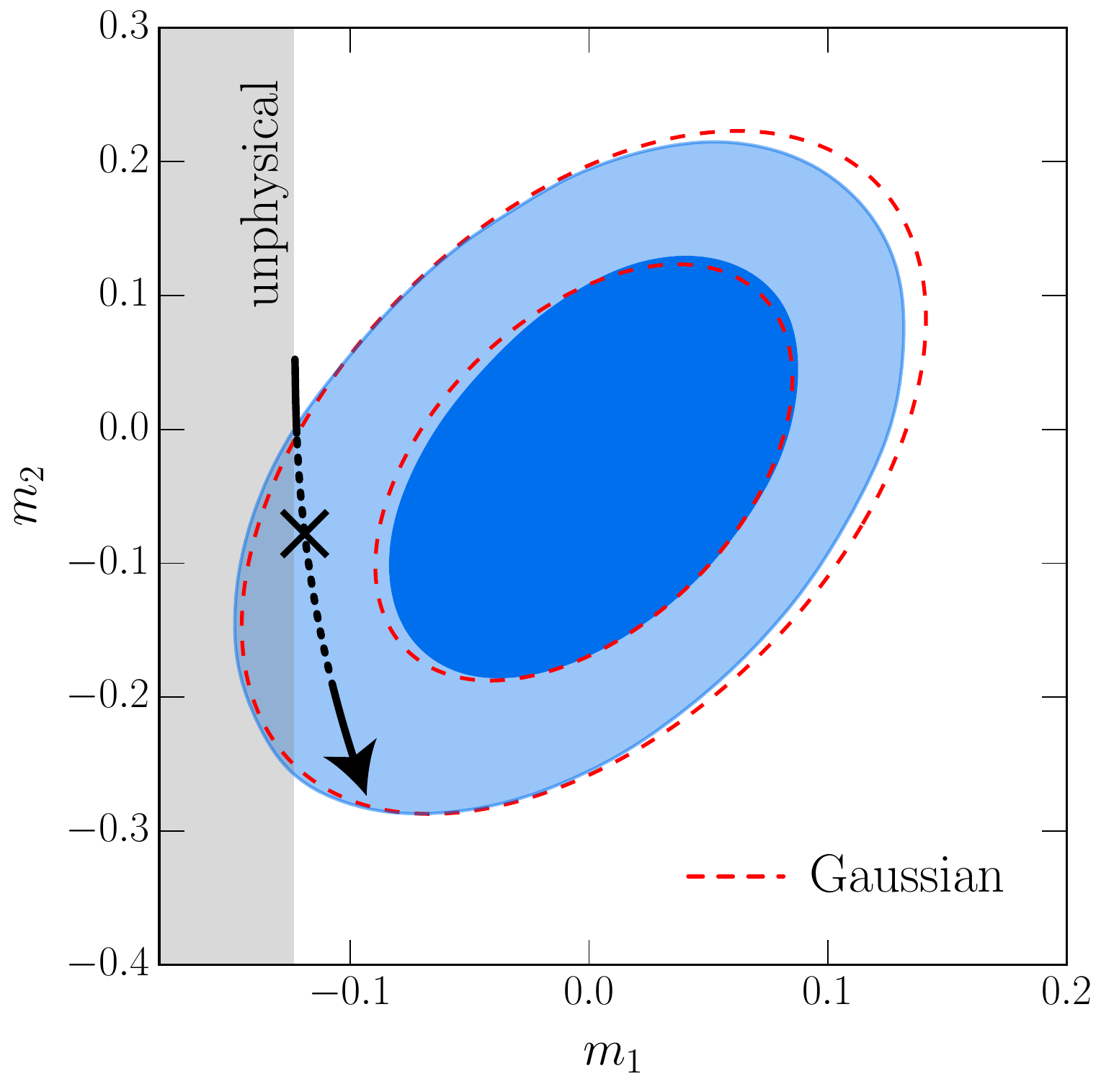}
             \caption{\footnotesize Best constrained PC plane, $m_1$-$m_2$ (magnified from Fig.~\ref{fig:mj_triangle_planck2015}). Tanh trajectories pass near the  boundary of the unphysical
             region (gray shaded) at low $\tau$ since reionization must be complete by $z=6$. 
             Also shown is the multivariate Gaussian approximation to the PC posterior using the means and covariance from Table~\ref{tab:PC_stats} [red dashed lines,
             68\% and 95\% CL, see Eq.~\ref{eq:gaussian}].
              }
            \label{fig:m1_m2_planck2015}   
\end{figure}

\section{Reionization Constraints}
\label{sec:MCMC}

We use the  Markov Chain Monte Carlo (MCMC) technique employing a modified version of COSMOMC\footnote{COSMOMC: \url{http://cosmologist.info/cosmomc}}~\cite{Lewis:2013hha, Lewis:2002ah} to sample from the posterior probability
density in the reionization and cosmological parameter space.    Our main analysis is on the
Planck 2015 data \cite{Aghanim:2015xee}  using the public likelihoods plik\_lite\_TTTEEE for high-$\ell$'s and lowTEB for low-$\ell$'s, which includes LFI but not HFI polarization.\footnote{We  have  tested that our results are robust to explicitly marginalizing foreground parameters, as opposed to using the premarginalized likelihood,  by separately running a PC MCMC using the lowTEB low-$\ell$ likelihood and the plikHM high-$\ell$ likelihood.}
For our cosmological parameters, we 
vary the base set of $\Lambda$CDM parameters (baryon density $\Omega_{b}h^2$,  cold dark matter density $\Omega_{c}h^2$,  effective acoustic scale $\theta_{\rm MCMC}$,  scalar power spectrum amplitude $\mathrm{ln}(10^{10} A_s)$ and tilt $n_s$).    We fix the neutrinos to their minimal 
contribution of one massive species with $m_{\nu}= 0.06$eV.   To those 
parameters we add the 5 PC mode amplitudes $m_1, \ldots, m_5$.
For comparison we also run a separate chain with the standard tanh ionization history, 
parametrized by the total reionization optical depth $\tau$.    We
assume flat priors in each of the given parameters.

In Fig.~\ref{fig:mj_triangle_planck2015}, we show the 1D and 2D marginalized posteriors in the 
PC amplitudes $m_a$ and for comparison the standard tanh ionization history 
with 68\% and 95\% ranges in its parameter $\tau$ projected onto the PCs.   Box bounds
represent the physicality prior from Eq.~(\ref{eq:individualprior}).   
In Table~\ref{tab:PC_stats} we give the corresponding means $\bar m_a$,  errors $\sigma(m_a)$ 
and  correlation matrix $R_{ab}$ which define the covariance matrix $C_{ab}$ as
\begin{equation}
C_{ab} = \sigma(m_a) \sigma(m_b) R_{ab}.
\label{eq:cov}
\end{equation}
Note that although the PCs are constructed to be uncorrelated for infinitesimal deviations from the fiducial
model, they do not remain so for finite deviations (see \cite{Mortonson:2008rx}).

\begin{table}[b]
\centering
\caption{PC chain means $\bar m_a$, standard deviations $\sigma(m_a)$, and correlation matrix $R_{ab}$. }
\label{tab:PC_stats}
\begin{tabular}{|r | r r@{\hskip 0.06in}|r r r r r|}
\hline
		
			  &  \multicolumn{1}{c}{$\bar m_a$} & \multicolumn{1}{c}{$\sigma(m_a)$}	 & \multicolumn{1}{|c}{$m_1$} & \multicolumn{1}{c}{$m_2$} & \multicolumn{1}{c}{$m_3$} & \multicolumn{1}{c}{$m_4$} & \multicolumn{1}{c|}{$m_5$} 
		\\ \hline
$m_1$ 
	& 0.002 & 0.053 & 1.000 & 0.450 & $-$0.432 & 0.273 & $-$0.073 \\ 
$m_2$ 
	& $-$0.030 &  0.101 & 0.450 & 1.000 & $-$0.262 & 0.055 & 0.072 \\ 
$m_3$ 
	& 0.019 &  0.128 & $-$0.432 & $-$0.262 &1.000 & $-$0.417 & 0.155 \\
$m_4$  
	& $-$0.012 & 0.143 &  0.273 & 0.055 & $-$0.417 & 1.000 & $-$0.428 \\ 
$m_5$ 
	& 0.026 & 0.143 & $-$0.073 & 0.072 & 0.155 & $-$0.428 & 1.000\\ \hline
\end{tabular}
\end{table}

Although 
all 5 PCs are measured to better than their physicality priors, unlike in the
WMAP5 analysis of Ref.~\cite{Mortonson:2008rx}, only $m_1$ and $m_2$ are bounded
substantially better.
In Fig.~\ref{fig:m1_m2_planck2015} we highlight this plane.   Note that in the
tanh model, low $m_1$ and high $m_2$, corresponds to small values of the total optical
depth and so these models skirt the edge of physicality there  given that the Universe must
be reionized by $z=6$.   In this MCMC, physicality priors are actually imposed after
the fact by eliminating samples in the shaded region.    Likewise the covariance matrix
in Eq.~(\ref{eq:cov}), is calculated before these samples are removed.   This ensures 
that the Gaussian approximation to the posterior is not distorted by the prior (see Eq.~\ref{eq:gaussian}).
As shown in Fig.~\ref{fig:m1_m2_planck2015}, the Gaussian approximation is fairly good
in the 68\% and 95\% CL regions of the $m_1$-$m_2$ plane, and we have checked that
it is equally good in the other planes.

\begin{figure}
          \includegraphics[width=0.9\linewidth]{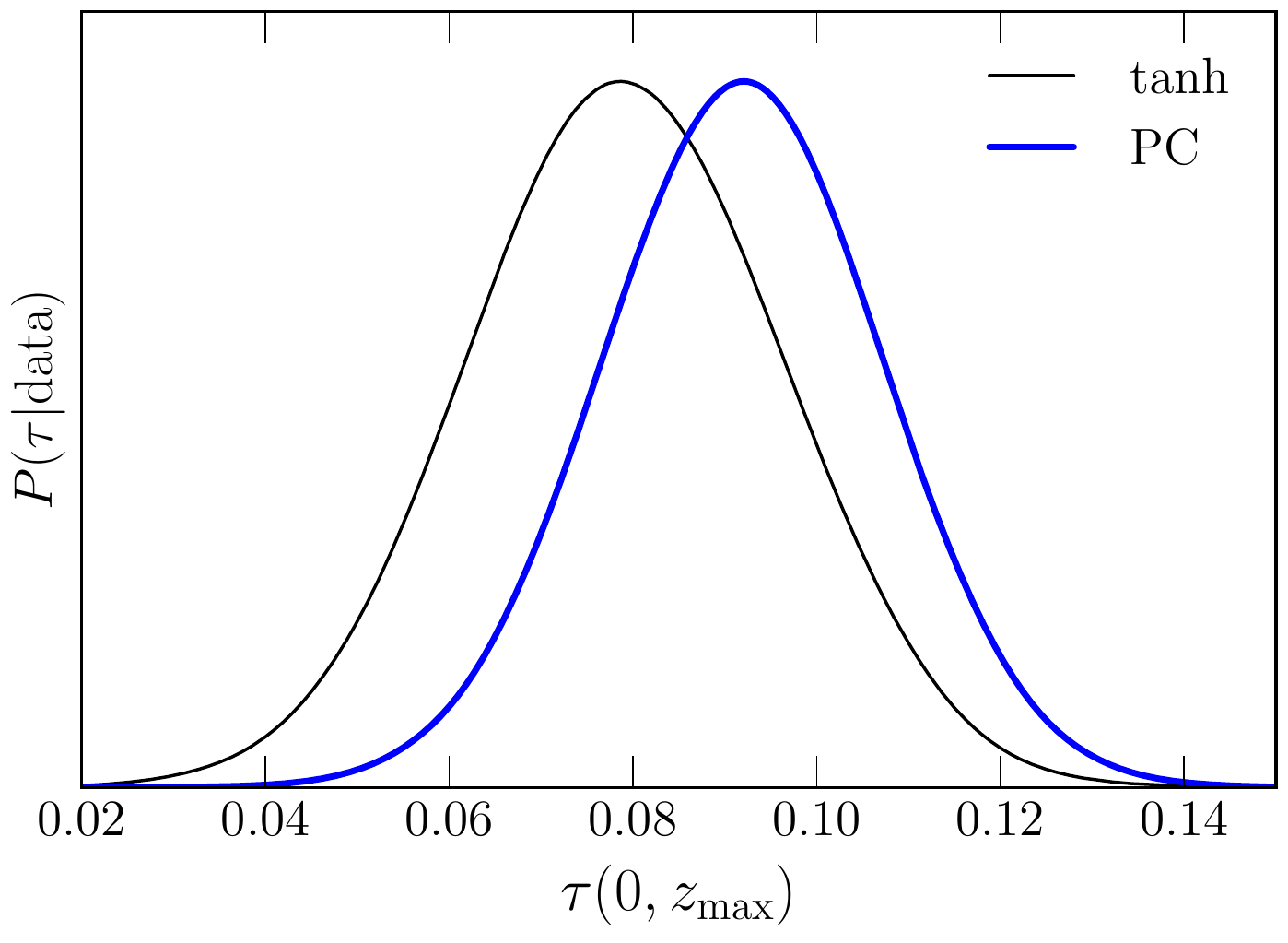}
             \caption{\footnotesize Posterior probability distribution of the total optical depth from the Planck 2015 tanh vs.~PC analyses.   The mean of the PC distribution is approximately 
             $1\sigma$ higher than the tanh distribution while the width remains comparable.
            }
            \label{fig:tau_posterior}   
\end{figure}

Interestingly, the standard tanh model is moderately disfavored in the wider, observationally complete,
PC space leading to changes in the inference for the optical depth  summarized in Table~\ref{tab:tau}. In Fig.~\ref{fig:tau_posterior}, we compare the posterior probability distributions for the total
$\tau(0,z_{\rm max})$ between the two chains.  In the PC analysis the total optical depth constraints shift up by
almost 1$\sigma$ from the tanh analysis while having comparable widths.   In $\Lambda$CDM the main consequence is that the amplitude of structure at $z=0$ goes from
$\sigma_8= 0.831 \pm 0.013$ in the tanh model to $\sigma_8=0.840 \pm 0.012$ in the PC analysis.

These changes are related to the fact that the
standard tanh models only skirt just inside the 2D 95\% CL even near the maximum likelihood
found in the chain (see Fig.~\ref{fig:m1_m2_planck2015}). The tanh ML model have the following parameters: $\Omega _{b}h^2 = 0.02224$, $\Omega_{c}h^2 = 0.1197$, $\theta_{\rm MCMC} =  1.04075$, $\mathrm{ln}(10^{10} A_s) =  3.0866$, $n_s = 0.9653$ and $\tau = 0.0765$.  Projected onto the PC space through Eq.~(\ref{eq:xetommu}), this value of $\tau$ corresponds to the $m_a$ parameter vector
\begin{equation}
\mathbf{m}^T =  \{ -0.119,  -0.078, 0.200,  -0.233, 0.129\}. 
\label{eq:tanhML}
\end{equation}
 Note that for the single best constrained component $m_1$, the tanh ML
model is more than $2\sigma$ off the mean.

\begin{figure}
          \includegraphics[width=1.0\linewidth]{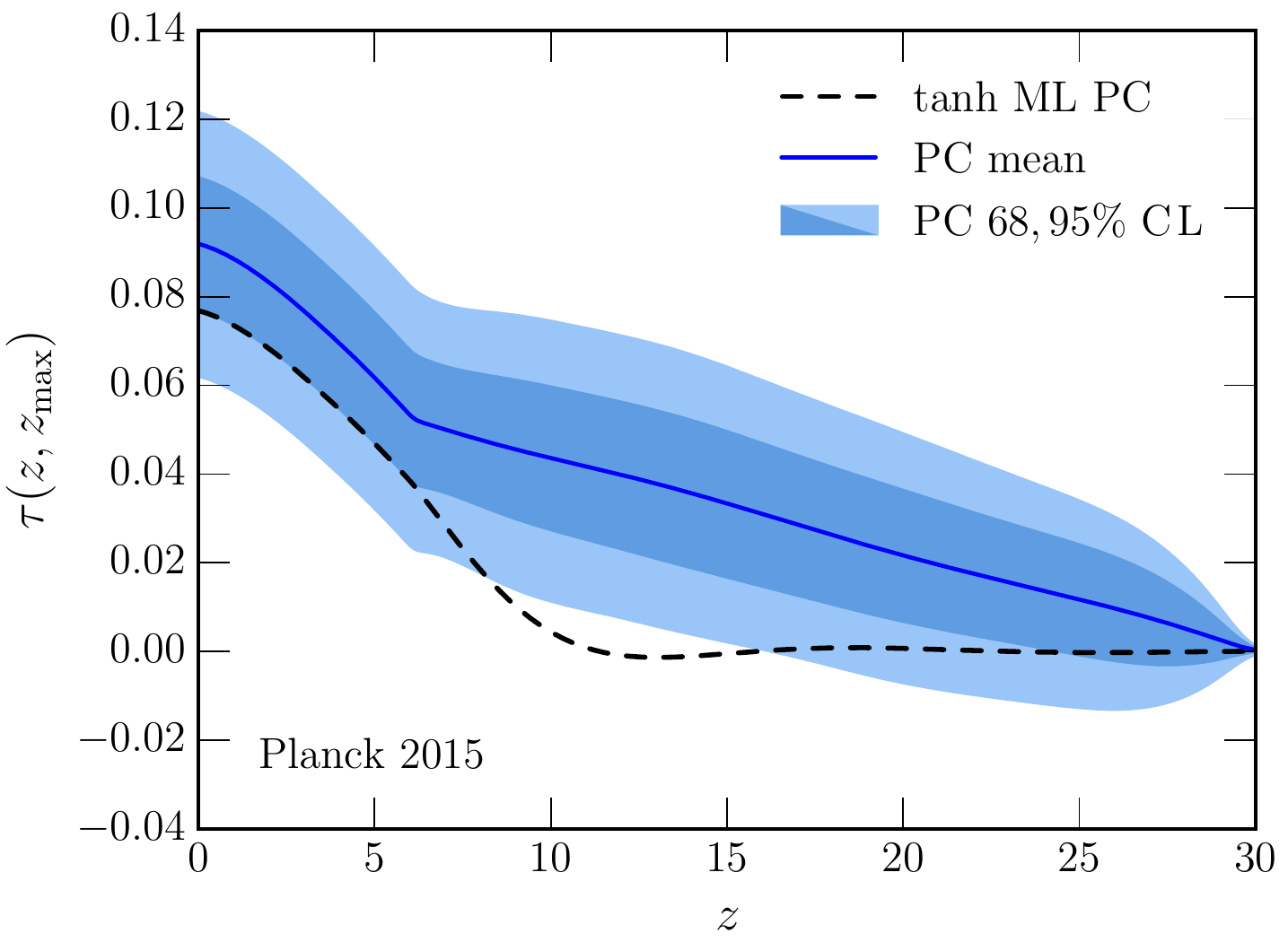}
             \caption{\footnotesize Cumulative optical depth $\tau(z,z_{\rm max})$ in the Planck 2015 analysis.  68\% and 95\% CL constraints from the complete PC analysis compared with the chain ML tanh model also constructed from PCs (dashed line).   PC analysis shows that the data prefer
             finite high redshift optical depth out to $z\gtrsim 16$ at 95\% CL.   The tanh model has essentially
             no optical depth for $z \gtrsim 10$ due to its steplike form. 
             Both model classes have the same ionization history for $z<6$.
         }
            \label{fig:tau_gtz_95cl}   
\end{figure}

In Fig.~\ref{fig:tau_gtz_95cl}, we explore the physical origin of this difference.
Here we show the $68\%$ and $95\%$ CL constraints on the cumulative optical
depth $\tau(z,z_{\rm max})$ as a parameter derived from the $m_a$ posterior probability.
The shape for $z<6$ is fixed by assumption.  
 Note that at $z\gg 1$, the cumulative optical depth from Eq.~(\ref{eq:cumtau}) becomes approximately
\begin{equation}
\tau(z,z_{\rm max}) \propto \frac{\Omega_b h^2(1-Y_p)}{(\Omega_m h^2)^{1/2}}
  \int_z^{z_{\rm max}} dz \, x_e(z) (1+z)^{1/2}  ,
\label{eq:cumtauapp}
\end{equation}
where $\Omega_m h^2$ parametrizes the sum over all of the nonrelativistic density components.  Integrals over
redshift in Eq.~(\ref{eq:cumtau}) can be precomputed for individual PCs once and for all in a fiducial cosmology
as in Fig.~\ref{fig:reion_basis} and then just summed with
$m_a$ weights and a rescaled prefactor following Eq.~(\ref{eq:cumtauapp}).

At the $95\%$ CL, the data
favor optical depth contributions at $z\gtrsim 16$.    For comparison, we
also plot the ML tanh model, projected onto the PC basis and calculated in the same way.   This model has essentially no optical depth contributions 
for $z>10$, not because the data forbid it but because of the functional form of the model.   More generally, because of its steplike form, the tanh family of models cannot generate high redshift
optical depth without also overproducing the total optical depth $\tau(0,z_{\rm max})$ 
(see Table~\ref{tab:tau}).

\begin{figure}
          \includegraphics[width=1.0\linewidth]{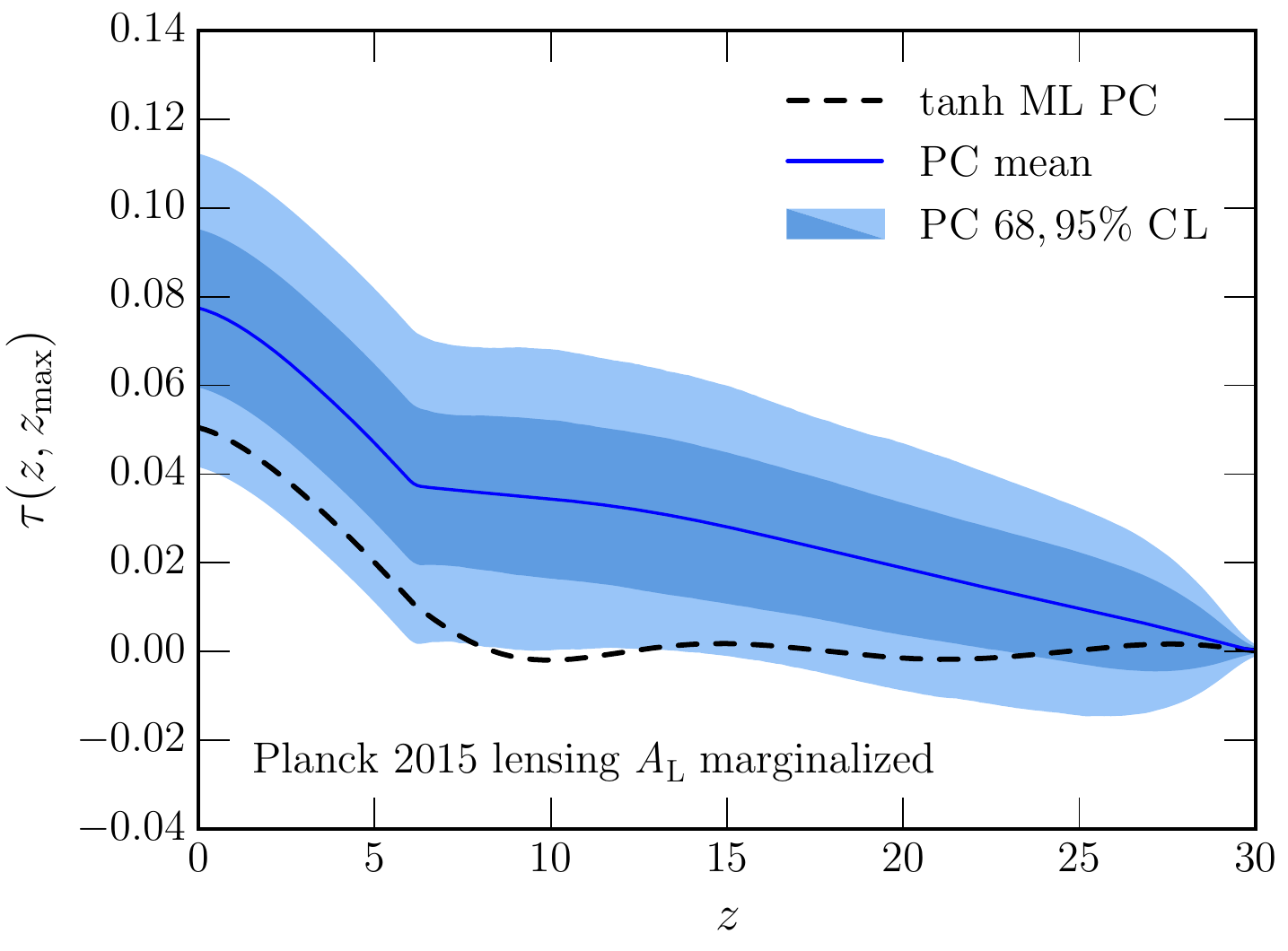}
             \caption{\footnotesize Cumulative optical depth $\tau(z,z_{\rm max})$ after marginalizing the lensing amplitude
             $A_L$ in the Planck 2015 analysis (as in Fig.~\ref{fig:tau_gtz_95cl} otherwise).  The total optical depth decreases for both the PC and tanh analyses 
             but mainly by lowering the low redshift contributions.   Preference for a finite high
             redshift contribution $\tau(15,z_{\rm max})$ remains at near the 95\% CL.  
             }
            \label{fig:tau_gtz_95cl_vary_al}   
\end{figure}

Given these differences, we explore further their origin in the data.    Constraints on the
optical depth are also affected by the temperature power spectrum indirectly and directly.  
Gravitational lens effects place constraints on the amplitude of the matter power spectrum through  $A_s$
and so in combination with  measurements of the temperature power spectrum which 
determine $A_s e^{-2\tau(0,z_{\rm max})}$
constrain $\tau(0,z_{\rm max})$ indirectly \cite{Hu:2001fb}.  
In particular, the 
Planck temperature power spectrum favors more gravitational lensing than is predicted
by the best fit $\Lambda$CDM parameters and hence tends to drive $\tau$ to larger
values \cite{Ade:2015xua}.

To test whether the preference for high redshift optical depth originates from gravitational
lensing and not large angle polarization, we follow the Planck 2015 analysis and marginalize
a multiplicative renormalization of the lens power spectrum $A_L$ by adding it to the 
parameter set of a new MCMC analysis of both the PC and the tanh model.  
In the PC case we obtain $A_L=1.11\pm 0.07$ and in the tanh case $A_L=1.15\pm 0.08.$  In Fig.~\ref{fig:tau_gtz_95cl_vary_al}, we show the
impact on $\tau(z,z_{\rm max})$.    As expected the total optical depth $\tau(0,z_{\rm max})$
is approximately 1$\sigma$ lower but notably the high redshift preference weakens only
moderately.   A finite $\tau(15,z_{\rm max})$ is still favored at nearly the 95\% CL (see also
Table~\ref{tab:tau}).
We conclude that much of the preference for high vs.~low redshift optical depth comes from
the large angle polarization data itself.

 \begin{figure}
          \includegraphics[width=1.0\linewidth]{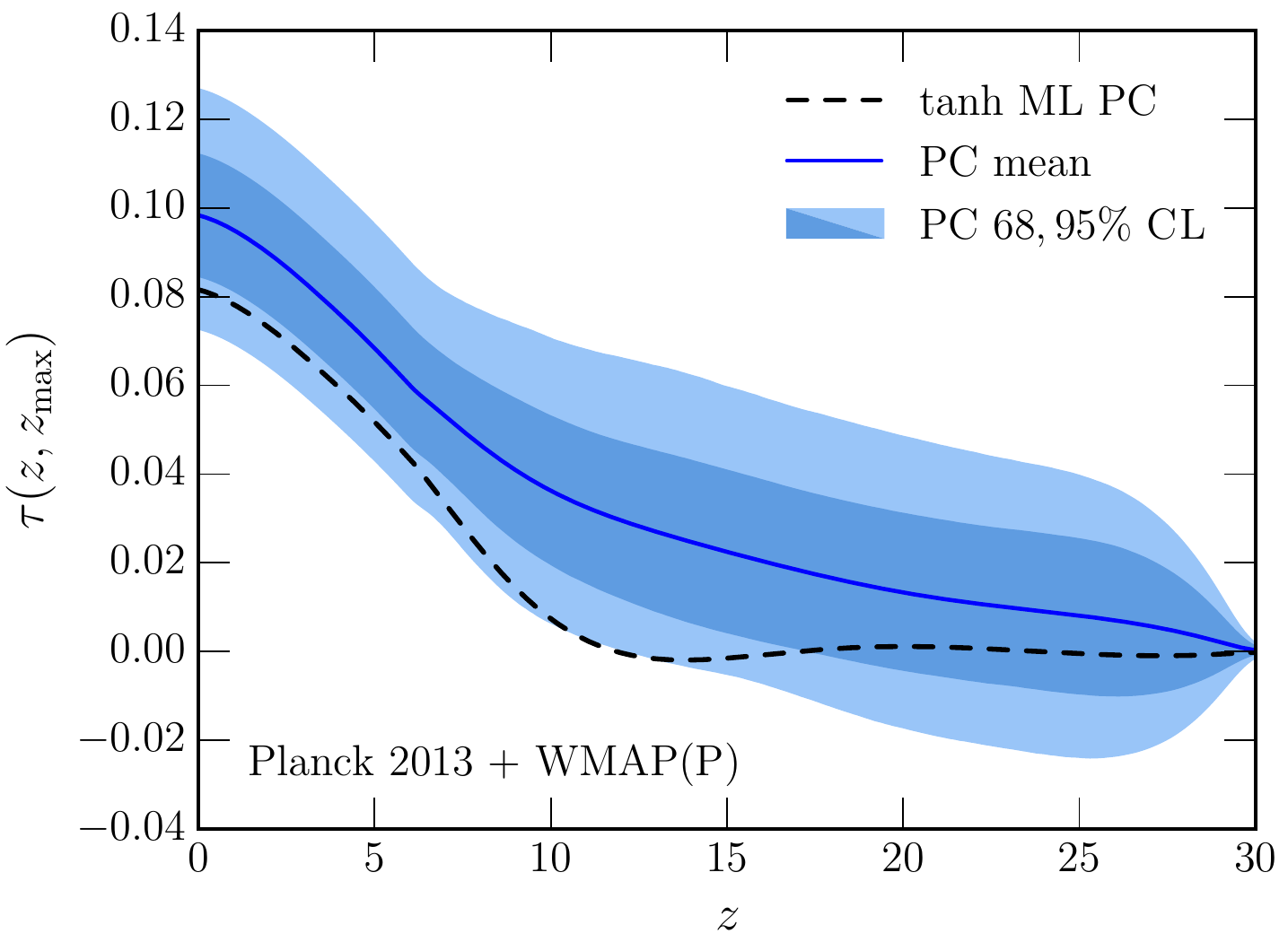}
             \caption{
            \footnotesize  Cumulative optical depth $\tau(z,z_{\rm max})$  in the Planck 2013 analysis with WMAP9
             polarization (as in Fig.~\ref{fig:tau_gtz_95cl} otherwise).  The total optical depth increases slightly for both the PC and tanh analyses while the high redshift contributions decrease.    Preference for a finite high
             redshift contribution $\tau(15,z_{\rm max})$ is lowered to closer to 68\% CL.   }
            \label{fig:tau_gtz_95cl_wmap_pol}   
\end{figure}

To further test this conclusion, we replace the Planck polarization data with WMAP9.   
In order to consistently analyze WMAP9 and Planck data sets with publicly available
likelihood codes, we also employ Planck 2013 instead of 2015 data.   Figure~\ref{fig:tau_gtz_95cl_wmap_pol} shows that this replacement has a  larger impact
on the high redshift end with the preference for $\tau(15,z_{\rm max})$ dropping to almost
$1\sigma$ (see also Table~\ref{tab:tau}).  Conversely, the constraints on $m_3,m_4,m_5$ remain largely the same
as the baseline Planck 2015 analysis indicating that they are constrained  mainly  by the temperature data.   As noted in Ref.~\cite{Mortonson:2008rx}
the Doppler effect from reionization imprints features on the temperature spectrum which
constrain the higher order PCs.
With the extended reach of the Planck temperature power spectrum as compared with
WMAP5, these features can be better separated from cosmological parameters.

\begin{figure}
          \includegraphics[width=0.95\linewidth]{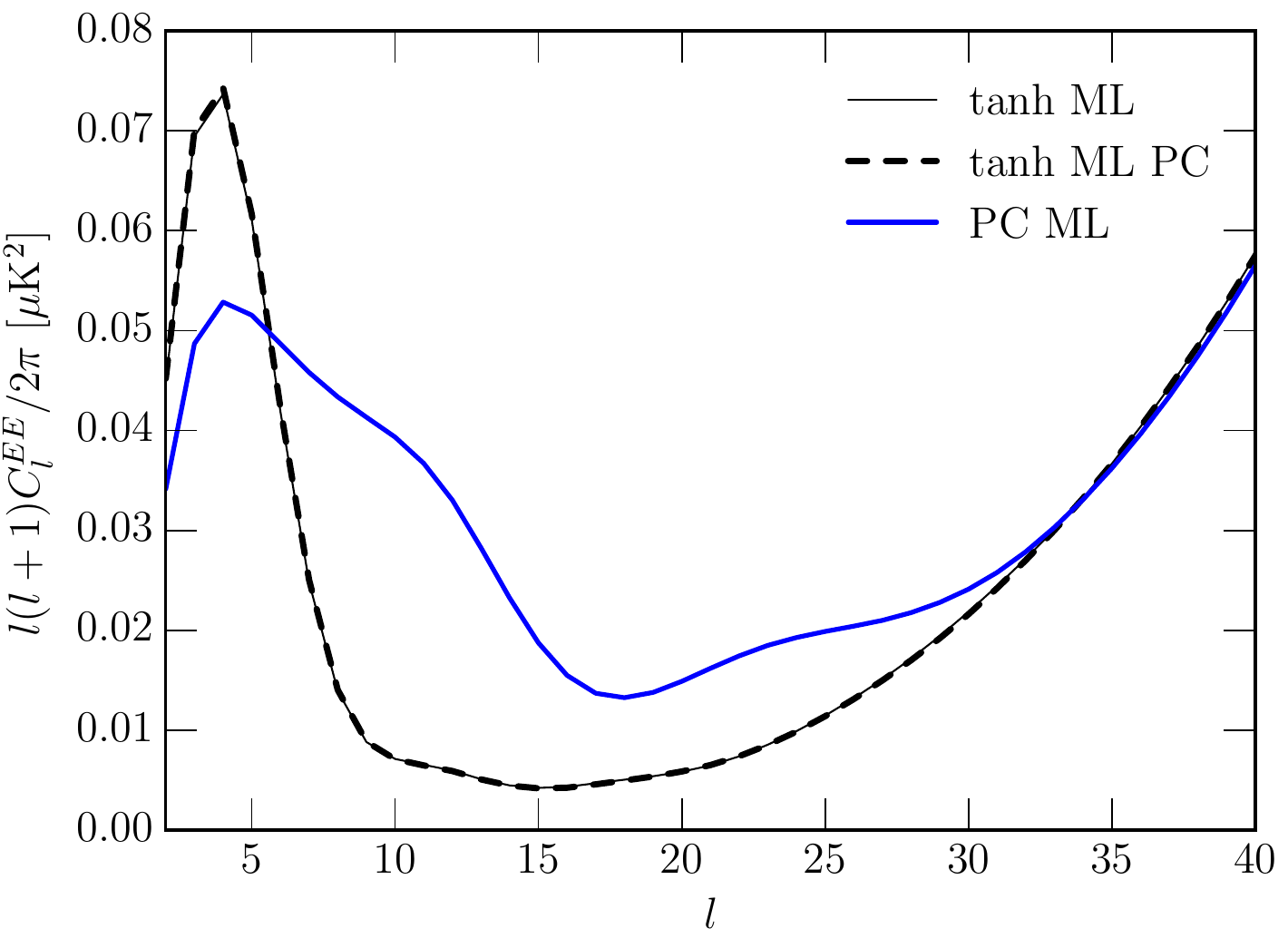}
             \caption{\footnotesize Polarization power spectrum $C_l^{EE}$ in the maximum likelihood models from the Planck 2015 PC and tanh chains.  Preference for high redshift optical depth in the PC analysis corresponds
             to a power spectrum with a broader reionization bump extending to higher multipoles than the
             tanh model.   To fit $\ell\lesssim 10$ data, tanh models predict much less power at
             $10 \lesssim \ell \lesssim 20$.
             Also shown is the tanh ML model calculated from the projection onto the 5 PC basis, showing its
             high accuracy and observational completeness.
             Note that the chain maximum is not necessarily the true maximum.
            }
            \label{fig:clee}   
\end{figure}

Since the preference for high redshift ionization mainly originates from the low multipole
polarization data, it is interesting to compare $C_\ell^{EE}$ for the maximum likelihood
tanh model from Eq.~(\ref{eq:tanhML}) to the maximum likelihood PC parameters.   The PC chain maximum has  $\Omega _{b}h^2 = 0.02230$, $\Omega_{c}h^2 = 0.1195$, $\theta_{\rm MCMC} =  1.04078$, $\mathrm{ln}(10^{10} A_s) =  3.1163$, $n_s = 0.9659$ and
\begin{equation}
\mathbf{m}^T=\{0.029,0.009,-0.026,0.085,0.062\}. 
\end{equation}
Note that in both cases the ML is simply the maximum found in the chain samples, not the true maximum in the parameter space.
Nonetheless, the difference in likelihood
of these two chain maximum models is $2\Delta\ln{\cal L}=5.3$ showing that the preference
is not just an artifact of parameter volume and priors.

In Fig.~\ref{fig:clee} we show that
the difference in $C_\ell^{EE}$ is that the PC ML has a much broader reionization bump that extends 
to higher multipoles.  This directly corresponds to the preference for high redshift ionization since the angular
scale of the feature is determined by the horizon at the redshift of scattering (see \cite{Hu:2003gh}, Fig. 2).
  In the tanh family of models, tight constraints at $\ell<10$ require
low power between $10 < \ell  < 30$ regardless of whether the data prefers it.

We also show in Fig.~\ref{fig:clee} that the projection of the tanh ML onto the 5 PC basis accurately
captures the reionization feature in the polarization spectrum with deviation of $\Delta C_\ell^{EE}/C_\ell^{EE} < 0.025$ for $\ell<40$ which tests the completeness of the 5 PC basis.
In fact most of the deviation is at $\ell \sim 10-20$ where the tanh model has minimal power so that the fractional accuracy still reflects a high absolute accuracy.

\begin{table}[]
\centering
\caption{Total  and high redshift optical depth constraints  for different model and data sets combinations.  Tanh models allow negligible high redshift optical depth whereas finite values are favored in the PC space.}
\label{tab:tau}
\begin{tabular}{| c  c |@{\hskip 0.06in} c@{\hskip 0.1in}  c|}
\hline Model  & Data            & $\tau(0,z_{\rm max})$             & $\tau(15, z_{\rm max})$ \\ \hline
PC  & P15         & 0.092 $\pm$ 0.015 & 0.033 $\pm$ 0.016                 \\ 
tanh  & P15       & 0.079 $\pm$ 0.017 & ...\\
PC $+A_L$    &P15  & 0.078 $\pm$ 0.018 & 0.028 $\pm$ 0.016                    \\ 
tanh $+A_L$    &P15  &   0.056 $\pm$  0.020  & ...         \\
PC & P13+WMAP(P) & 0.098 $\pm$ 0.014 & 0.022 $\pm$ 0.018                  \\ 
tanh & P13+WMAP(P) & 0.090 $\pm$ 0.013 &  ...           \\ \hline
\end{tabular}
\end{table}

\section{Ionization History Likelihood}
\label{sec:likelihood}

One of the main benefits of our PC reionization analysis is that it completely encapsulates
the information from the large angle polarization measurements on the ionization history within the
given redshift range.
With this single analysis, it is possible to infer constraints on any
given reionization model in the same range.   In this section, we provide a concrete prescription for an
effective likelihood function for the data given an $x_e(z)$ model.

The MCMC PC analysis of the previous section returns the posterior probability density in 
the space of PC
amplitudes $m_a$ given the data and flat priors on the parameters, marginalized over 
cosmological parameters.   Invoking Bayes' theorem,
we can reinterpret it as an effective likelihood of the data given $m_a$.   To determine
the likelihood given  $x_e(z)$ instead, we simply need to project it onto the PC  basis using
Eq.~(\ref{eq:xetommu}).

In practice, the MCMC only provides a sample of the PC posterior, represented by
discrete elements in the chain with parameter values $\mathbf{m}_i = \{m_1, \ldots, m_5\}$
and multiplicities $w_i$,
whereas a given model produces a continuum of values for $\mathbf{m}$.   
Given the samples, we approximate the effective likelihood with a kernel
density estimator of the form 
\beq
{\cal L}_{\rm PC}\left({\rm data}|\mathbf{m} \right)  = \sum_{i = 1}^{N} w_i K_f(\mathbf{m}-\mathbf{m}_i),
\eeq
where $N$ is the total number of elements in the chain and the overall normalization is arbitrary. Here $K_f$ is a smoothing kernel that
makes the function estimate continuous at the expense of artificially broadening the distribution.
We choose the shape of $K_f$ to be a multivariate Gaussian of zero mean and 
covariance $f\mathbf{C}$ where $\mathbf{C}$ is the $m_a$ 
covariance matrix estimated from the
chain from Table~\ref{tab:PC_stats} and Eq.~(\ref{eq:cov}).    For a Gaussian posterior, the effect of smoothing
is to increase the covariance by $1+f$ or the errors by approximately $1+f/2$.   
To minimize the amount of smoothing required to capture the behavior of models
in the tail of the distribution like the standard tanh model, we oversample the posterior by running  the 
chain past normal convergence requirements for a total of $N\approx 1.4 \times 10^6$ chain elements.
In practice we choose $f=0.14$.  Note also that we employ the full chain without physicality priors since the smoothing kernel transfers information across these boundaries.

\begin{figure}
          \includegraphics[width=0.7\linewidth]{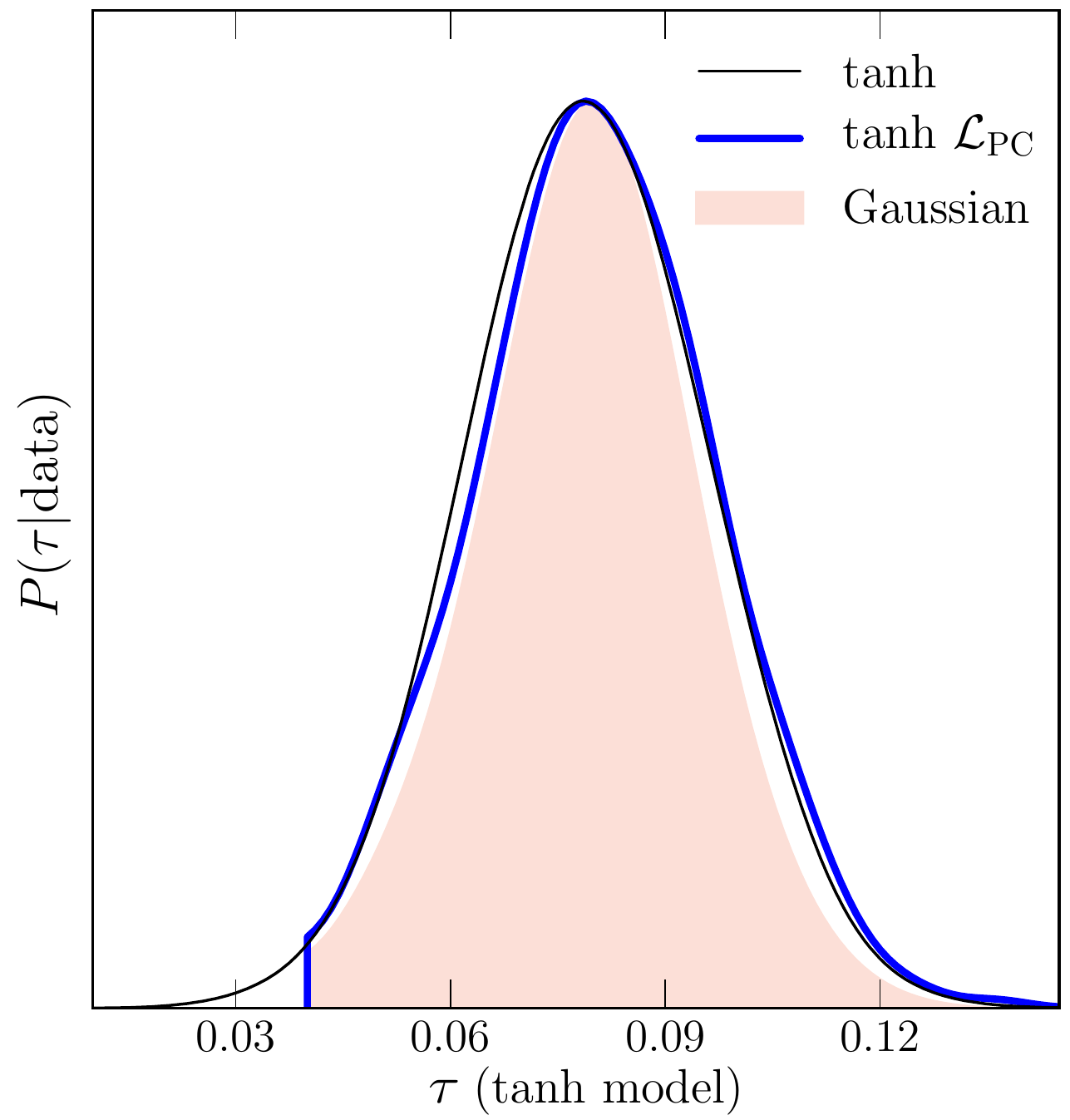}
             \caption{\footnotesize Effective likelihood analysis (${\cal L}_{\rm PC}$ thick blue) on the tanh model compared with direct constraints (thin black).  Even though the tanh model lives in the tails of the PC posterior, the effective likelihood models the direct constraint to a fraction of its width.   
             Also shown is the Gaussian approximation to the effective likelihood (shaded region). 
             The distributions are truncated at low optical depth since the fact that the Universe must be reionized by $z=6$ is enforced by the PC decomposition. }
            \label{fig:tanh_tau_posterior}   
\end{figure}

To illustrate and test this approach, we  use the effective likelihood to compare constraints on
the 
standard tanh model. In this case the model parameter is the total $\tau$ 
given the tanh ionization history $x_e(z; \tau)$.   We can then construct the posterior probability of
$\tau$ as usual via Bayes' theorem and the effective likelihood given $\mathbf{m}(\tau)$
\begin{equation}
P(\tau| {\rm data}) \propto {\cal L}_{\rm PC}\left[ {\rm data}|\mathbf{m}(\tau) \right] P(\tau).
\end{equation}
To match the MCMC analysis of the standard tanh model we take flat priors $P(\tau)=1$.   
For the conversion between the ionization history and $\tau$ we take the cosmological parameters
of the ML tanh model.  

In Fig.~\ref{fig:tanh_tau_posterior}, we compare the posterior probabilities from the direct MCMC
analysis and the effective PC likelihood.   The distributions match to much better than $1\sigma$
in their means and widths.    This is a fairly stringent test on the method given that 
tanh models live in the tails of the PC posterior.  
 The cutoff at low $\tau$ in the effective likelihood method
 simply reflects the $z_{\rm min}$ restriction for the PCs which assumes hydrogen ionization occurred at $z>6$ as is observationally the case.  
 
It is also interesting to compare these results to an even simpler effective likelihood.  In
Fig.~\ref{fig:tanh_tau_posterior} (shaded region), we also show the result of approximating
 the $m_a$ posterior as a multivariate Gaussian with mean $\bar m_a$
 and covariance $\mathbf{C}$
 \begin{equation}
 {\cal L}_{\rm G}\left({\rm data}|\mathbf{m} \right) = \frac{ e^{-\frac{1}{2} ({\bf m}-\bar{\bf m})^T {\bf C}^{-1} ({\bf m}-\bar{\bf m}) } }{\sqrt{(2\pi)^5 | \mathbf{C}| }}
.
 \label{eq:gaussian}
 \end{equation}
 For an extremely fast but approximate effective likelihood and for models near the peak
 of the distribution, the Gaussian approximation may suffice.
 
  Validated on the tanh model, our effective likelihood technique allows a rapid exploration of other models without the need for
 a separate MCMC analysis.
 To illustrate this usage, we consider the power-law (PL) models that the Planck Collaboration also analyzed for the HFI data~\cite{Adam:2016hgk}.   The
 PL model 
  \begin{equation}
\frac{ x_{\rm e}(z)}{ 1 + f_{\rm{He}}} =  
  \begin{cases}
    1, & 6< z<z_{\rm end}, \\
  \left(\frac{z_{\rm early}-z}{z_{\rm early}-z_{\rm end}}\right)^\alpha, 
      &z_{\rm end}\le z\le z_{\rm early}, \\
      0, & z_{\rm early}< z < z_{\rm max}, \\
  \end{cases}
  \label{eq:power_law}
\end{equation}
 allows for  an extended and asymmetric ionization history unlike the tanh model.  
 Here $z_{\rm early} < z_{\rm max}$ is a  parameter that truncates reionization 
 at $z_{\rm early}$ but has little effect at lower redshift.  
 In order to mimic the analysis in Ref~\cite{Adam:2016hgk} but allow for the weaker
 constraints from Planck 2015, we fix $z_{\rm early}=23.1$ so that it does not significantly
 impact the analysis.
 The power-law index $\alpha$ controls the duration of reionization $\Delta z$ 
 defined so that $x_e(z_{\rm end}+\Delta z) = 0.1 (1 + f_{\rm He})$.  It is therefore
 more convenient to parametrize the model with $z_{\rm end}$ and $\Delta z$.  We employ the effective PC likelihood approach with flat priors for this two
 parameter family within their allowed ranges.
 
\begin{figure}
          \includegraphics[width=0.75\linewidth]{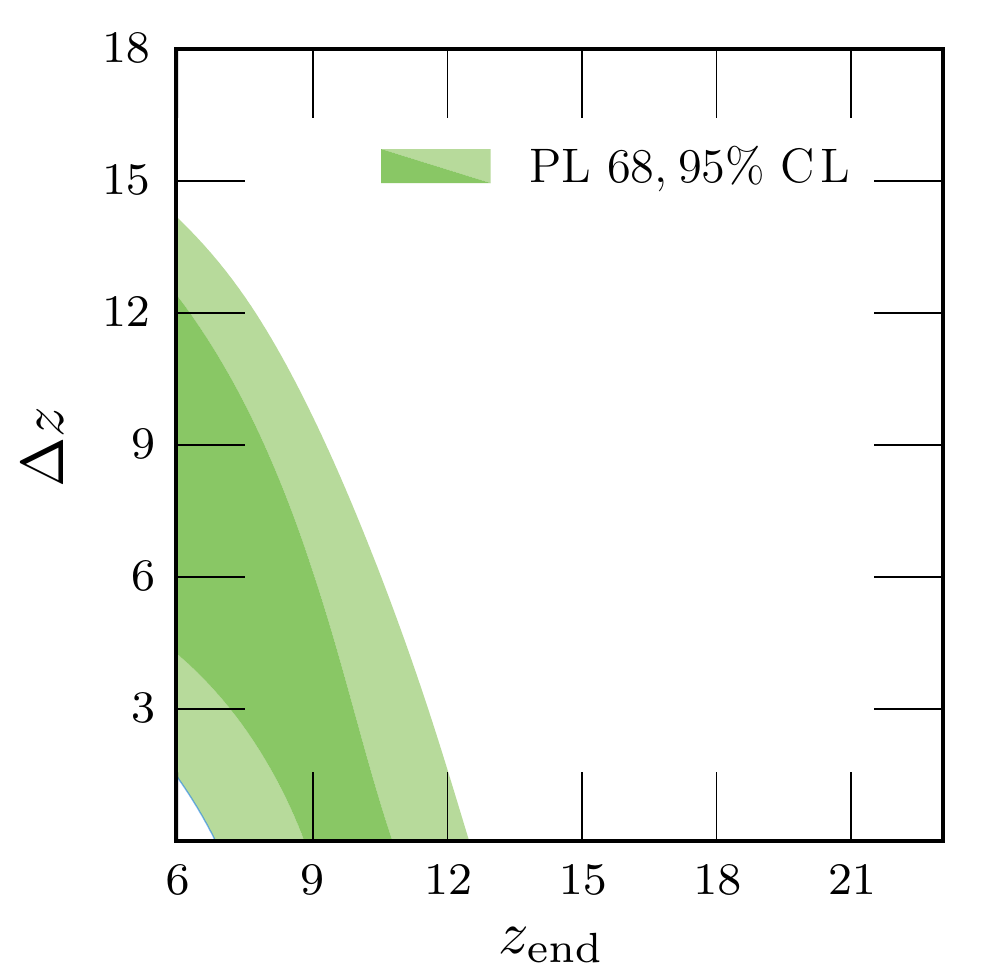}
             \caption{\footnotesize Effective likelihood constraints on the two parameters of the power-law (PL) model:  end ($z_{\mathrm{end}}$) and duration ($\Delta z$) of reionization in redshift.  The PL functional form forces the two parameters to be anticorrelated given
              the constraint on the total optical depth and hence does not allow a separate
              component of high-$z$ ionization. 
           }
            \label{fig:power_law_2D}   
\end{figure}

\begin{figure}
          \includegraphics[width=1.0\linewidth]{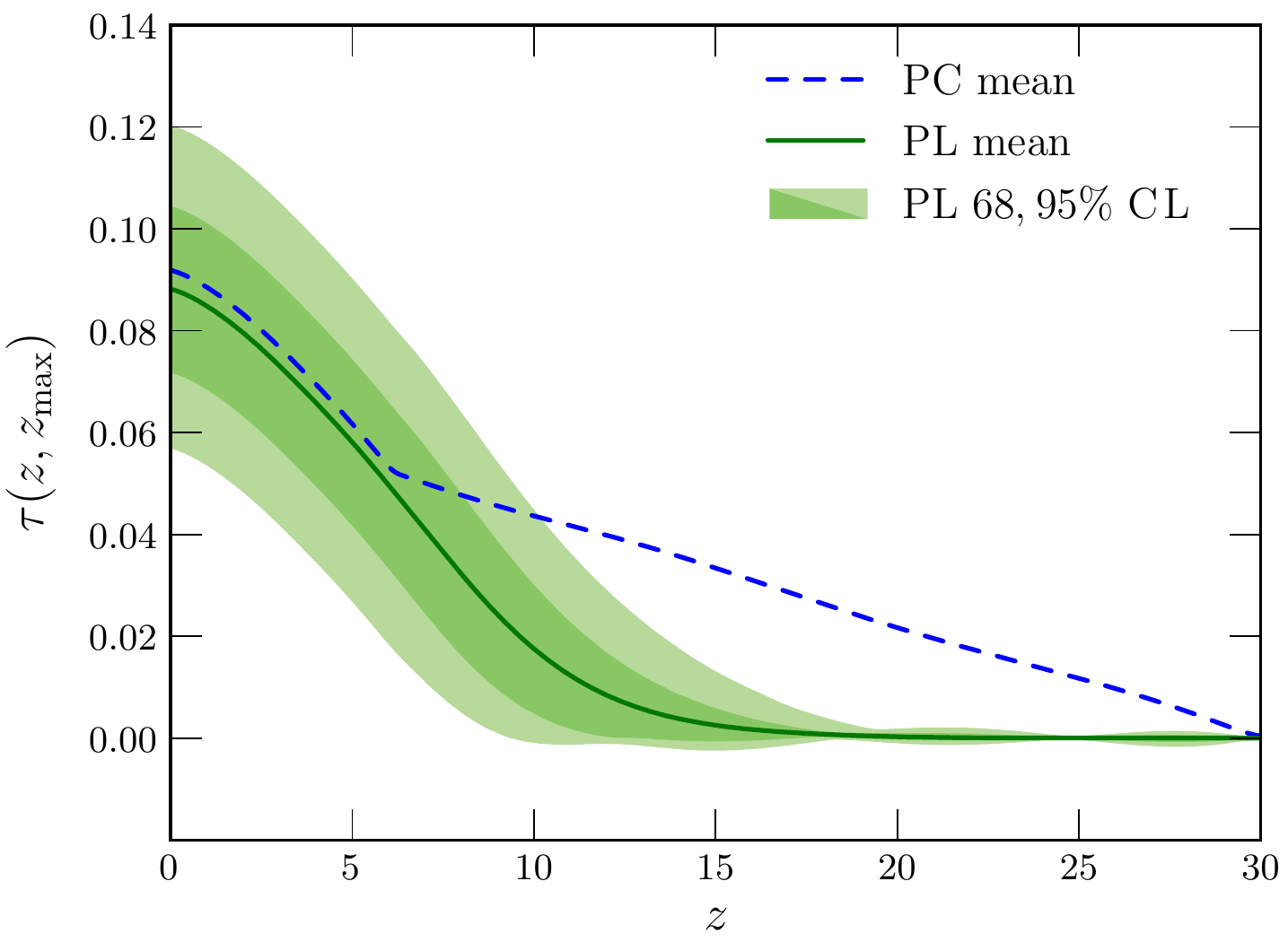}
             \caption{\footnotesize Effective  likelihood constraints on the cumulative optical depth $\tau(z,z_{\rm max})$ for the PL model.   Although the PL model allows an adjustable duration of reionization, its functional form does not permit the high redshift optical depth
             of the PC mean (blue dashed) without violating the total optical depth constraint.  }
            \label{fig:tau_gtz_power_law}   
\end{figure}

 In Fig.~\ref{fig:power_law_2D}, we show the 2D posteriors in these parameters.   
 The Planck 2015 data do indeed allow an extended period of reionization if $z_{\rm end}
 \sim 6$ but do not particularly favor it over a prompt reionization that maintains the
 same total optical depth.    Like the tanh model, the PL model links the low redshift and high
 redshift ionization history by assumption of a functional form appropriate for models with
 a single phase of reionization.  
 One should
therefore not equate a constraint on the duration of reionization in the PL model with
a constraint on high redshift ionization in a more general context. 
 To see this more quantitatively,
in Fig.~\ref{fig:tau_gtz_power_law} we show the  cumulative optical depth $\tau(z,z_{\rm max})$ constraints in the PL context. The PL models allow very little optical depth at $z > 15$
compared with the mean in the full PC space from Fig.~\ref{fig:tau_gtz_95cl}.  

The PL example also illustrates the fact that to encompass the region favored by the PC
effective likelihood, a reionization model would need to have an additional source of high redshift
 ionization that is not directly linked in functional form to its low redshift behavior.    We explore such
 cases in a separate work \cite{Miranda:2016trf}.  In fact,  we also 
 expect the effective likelihood method to work even better for models that are 
 favored by the data since the underlying MCMC sample better represents these models.
  
\section{Discussion}
\label{sec:discussion}
By analyzing the Planck 2015 data with an observationally complete PC basis
for the ionization history, we show that it
    allows and even favors high redshift, $z\gtrsim 15$, optical depth at the $\sim 2\sigma$ level.
    The standard analysis which includes just the total optical depth and assumes a sharp
    steplike transition excludes this possibility by prior assumption of form rather than
    because it is required by the data.  The same is true for power-law models that additionally vary the duration of reionization.
    
While a $2\sigma$ result amongst the 5 PC parameters is not on its own surprising,
it originates from the first and best constrained component and hence has consequences for the total optical depth.   The total optical depth is important for understanding a  host of
other cosmological parameters from the amplitude of the current matter power spectrum
$\sigma_8$ to the inferences from CMB lensing.   At the very least, this analysis
highlights the need for a complete treatment of CMB reionization observables to guarantee a robust interpretation of the optical depth. 

This preference for extra high redshift optical depth mainly originates from the large angle polarization spectrum in the
Planck 2015 data and appears related to excess power in the multipole range
$10 \lesssim \ell \lesssim 20$.   It is only slightly weakened by marginalizing gravitational lensing information in the temperature power spectrum, which is known to favor a higher optical depth, but more significantly changed
by replacing the Planck LFI  with WMAP9 polarization data.   

While excess polarization power in this range favors additional sources of high redshift ionization
such as population III stars or dark matter annihilation it could also indicate contamination
from systematics and foregrounds.    The latter have been significantly improved in the
as yet proprietary Planck 2016 intermediate results.  These results indicate that the low redshift end of the optical 
depth as tested by steplike models,
or equivalently the low $\ell$ polarization power, is both better measured and lower than the central value in the Planck 2015
data \cite{Aghanim:2016yuo}.  On the other hand, these results exacerbate the tension
with gravitational lensing in the shape of the temperature power spectrum
which probes the total optical depth.   It will be interesting to see if this complete analysis
still prefers an additional high redshift component in the final Planck release.

Regardless of the outcome of resolving the mild tension between steplike reionization 
scenarios and the Planck 2015 data,  the complete PC approach developed here is useful
because with a single analysis one can infer constraints on the parameters of any 
reionization model within the specified redshift range, here $6 < z <30$, but easily extensible to any desired range.
What has presented an obstacle for this approach in the past is the lack of tools for converting
posterior parameter constraints on the PCs to parameter constraints on models and so 
be able to combine them with other sources of reionization information. 
For example, the ionization history can also be tested in the CMB through the kinetic Sunyaev-Zel'dovich effect from temperature
fluctuations beyond the damping scale, but in a manner that is highly model dependent
(e.g. \cite{Mortonson:2010mi,Zahn:2011vp,Battaglia:2012im,Adam:2016hgk}).

Towards this end, we have developed and tested  an effective likelihood code
for inferring constraints on any given ionization history provided by a model.   This approach
should be especially useful in constraining models where small high redshift contributions to the
optical depth need to be separated from the total.

 \medskip
 \noindent {\it Acknowledgments}:   We thank Austin Joyce, Adam Lidz, and Pavel Motloch for
 useful discussions.  W.H. thanks the Aspen Center for Physics, which is supported by National Science Foundation Grant PHY-1066293, 
where part of this work was completed. C.H. and W.H. were supported by NASA ATP NNX15AK22G, U.S.~Dept.\ of Energy Contract No. DE-FG02-13ER41958, and  the Kavli Institute for Cosmological Physics
at the University of Chicago through Grants No. NSF PHY-0114422
and No. NSF PHY-0551142. Computing resources were provided by the University of Chicago Research Computing Center.  V.M. was supported in part by the Charles E.~Kaufman Foundation, a supporting organization of the Pittsburgh Foundation.

\vfill

\bibliography{ReiP15}

\end{document}